\documentclass{appolb}
\usepackage{cite,./mcite}   
\usepackage{graphicx}
\usepackage{amsmath}

\begin{document}
\title{A parton branching  algorithm with transverse momentum dependent splitting functions%
\thanks{Presented at ``Diffraction and Low-$x$ 2022'', Corigliano Calabro (Italy), September 24-30, 2022.}%
}
\author{Aleksandra Lelek
\address{University of Antwerp, Belgium}
}
\maketitle
\begin{abstract}

Parton branching methods underlie the Monte Carlo (MC) generators, being therefore of  key importance for obtaining high energy physics predictions.
We construct a new parton branching algorithm which for the first time  incorporates the off-shell, transverse-momentum dependent (TMD) splitting functions,   defined from the high-energy limit of partonic decay amplitudes. 
Based on these TMD splitting functions we construct a new  TMD Sudakov form factor.
We present the first MC implementation of the algorithm for the  evolution of the TMD and integrated parton distribution functions (PDFs). We use this implementation to evaluate small-$x$ corrections to the distributions and to verify the momentum sum rule.
The presented study is a first step towards a full TMD MC  generator covering the small-$x$ phase space.
\end{abstract}
  
\section{Introduction}

The outcomes of high energy collider experiments depend to a large extent on event simulations obtained with MC generators. So do the planning and development of future machines and measurements   \cite{Azzi:2019yne,Feng:2022inv,Mangano:2016jyj,LHeC:2020van,Proceedings:2020eah}. The baseline MCs are based on the description of hadron structure  provided by collinear PDFs \cite{Kovarik:2019xvh}, while a more complete, 3D description of hadron structure is given by TMD PDFs \cite{Angeles-Martinez:2015sea}. There are thus efforts to include elements of TMD physics in the modern MC generators and in the parton-branching algorithms on which they are based. The idea of the work \cite{Hautmann:2022xuc} described in this article is to include the TMD splitting functions obtained from the high-energy (or small-x) limit of partonic amplitudes \cite{Catani:1994sq} in a parton branching algorithm, with the goal to incorporate in the parton evolution both small-x and Sudakov contributions. Thanks to its applicability over a wide kinematic region, the algorithm provided by the TMD Parton Branching (PB) method \cite{Hautmann:2017xtx,Hautmann:2017fcj} was chosen to perform this research.

\section{The TMD Parton Branching method}

The PB method is a flexible, widely applicable MC approach to obtain QCD high energy predictions based on TMD PDFs,  simply called TMDs. 
One of its main ingredients is a forward evolution equation \cite{Hautmann:2017xtx,Hautmann:2017fcj}. 
The evolution of the parton density is expressed in terms of real, resolvable branchings and virtual and non-resolvable contributions, which are treated with Sudakov form factors. 
 Thanks to the momentum sum rule \footnote{Momentum sum rule for the DGLAP splitting functions $P_{ab}(z,\mu^2)$ yields $\sum_a\int_0^1 \textrm{d} z \; z P_{ab}(z,\mu^2) = 0$. }
 and unitarity, the  Sudakov form factor can be written in terms of real, resolvable splittings and interpreted as a non-emission probability.    
Owing to the simple, intuitive picture of the evolution in terms of cascade of branchings and the probabilistic interpretation of the splitting functions and the Sudakov form factors, the PB evolution equation can be solved with MC techniques using a parton branching algorithm.

Additionally to the evolution equation, PB provides also a procedure to fit parameters of the initial distribution to the experimental data using \texttt{xFitter} platform  \cite{Alekhin:2014irh}. Obtained PB TMDs and PDFs \cite{BermudezMartinez:2018fsv,Jung:2021vym,Jung:2021mox}  are  accesible via TMDlib \cite{Abdulov:2021ivr} and  in LHAPDF \cite{Buckley:2014ana} format for the usage  in (TMD) MC generators.  A generator of a special importance  is the TMD MC generator Cascade \cite{Baranov:2021uol} where 
    the  TMD initial state parton shower is implemented  with the backward evolution guided by the PB TMDs. 
The PB method provides the  procedure to match PB TMDs with next-to-leading order (NLO) matrix elements \cite{BermudezMartinez:2019anj} to obtain predictions. Recently, there was also a merging procedure developed \cite{BermudezMartinez:2021lxz}. 
The PB method was used to study different  evolution scenarios 
like ordering conditions or resolution scales, see e.g. \cite{Hautmann:2017xtx,Hautmann:2019biw}. The PB predictions have been calculated for multiple measurements, in very different energy and mass regimes, including hadron colliders, fixed target experiments and $ep$ collider  \cite{BermudezMartinez:2018fsv,BermudezMartinez:2019anj,BermudezMartinez:2020tys,Yang:2022qgk,Abdulhamid:2021xtt,H1:2021wkz}.

All those successful PB studies were performed with  the  DGLAP \cite{Gribov:1972ri,Lipatov:1974qm,Altarelli:1977zs,Dokshitzer:1977sg} splitting functions  calculated in the collinear approximation. However, in some infrared-sensitive phase space regions, the collinear approximation is not enough
\cite{Dooling:2012uw,Dooling:2014kia}. In this work the PB approach was extended by using the TMD splitting functions \cite{Catani:1994sq,Gituliar:2015agu,Hentschinski:2016wya,Hentschinski:2017ayz}.

\section{TMD splitting functions}
The concept of the TMD splitting functions originates from the high energy factorization \cite{Catani:1994sq}, where the TMD splitting function for the splitting of an off-shell gluon into quark, $\widetilde{P}_{qg}$, was calculated. The other channels were obtained in \cite{Gituliar:2015agu,Hentschinski:2016wya,Hentschinski:2017ayz}. 
The splitting functions have well defined collinear and high energy limits. 
It was demonstrated that in the limit of small incoming transverse momenta, after angular average, the TMD splitting functions converge to the  DGLAP leading order (LO) splitting functions. For finite transverse  momenta, the TMD splitting function \cite{Catani:1994sq} can be written as an expansion in powers of the transverse momenta with  $z$-dependent coefficients, which, after convoluting them with the TMD gluon Green's functions \cite{Kuraev:1977fs,Balitsky:1978ic}, give the
 corrections to the  splitting function logarithmically enhanced for $z\rightarrow 0$. Therefore, the work presented next on the implementation of 
TMD splitting functions in the PB method can be viewed as a step toward 
constructing full MC generators for small-$x$ physics  (see e.g.  \cite{Chachamis:2015zzp,Andersen:2011zd,Jung:2010si,Hoeche:2007hlb,Golec-Biernat:2007tjf}).

\section{TMD splitting functions in the PB method}

The DGLAP splitting functions $P_{ab}^R (z, \mu^{\prime})$ were replaced by the TMD ones $\tilde{P}_{ab}^{R}\left(z, k_{\bot} +(1-z)\mu_{\bot}^{\prime}, \mu_{\bot}^{\prime}\right)$ in the PB evolution equation  for the momentum weighted parton density $x{\mathcal{A}}_a = \tilde{\mathcal{A}}_a$  \cite{Hautmann:2017fcj}
\begin{multline}
\tilde{\mathcal{A}}_a\left( x,k_{\bot}^2, \mu^2\right) = 
 \Delta_a\left(\mu^2,k_{\bot}^2\right)\tilde{\mathcal{A}}_a\left( x,k_{\bot}^2, \mu_0^2\right) + 
 \sum_b\int\frac{d^2\mu_{\bot}^{\prime}}{\pi\mu_{\bot}^{\prime 2}}\Theta(\mu_{\bot}^{\prime 2}-\mu_0^2)\Theta(\mu^2-\mu_{\bot}^{\prime 2})
\\
\times  \int\limits_x^{z_M }\textrm{d}z\,  \frac{ \Delta_a\left(\mu^2, k_{\bot}^2  \right)  }  
  { \Delta_a\left(\mu_{\bot}^{\prime 2}, k_{\bot}^2 \right)}  \tilde{P}_{ab}^{R}\left(z, k_{\bot} +(1-z)\mu_{\bot}^{\prime}, \mu_{\bot}^{\prime}\right) 
  \tilde{\mathcal{A}}_b\left( \frac{x}{z},  (k_{\bot}+(1-z)\mu_{\bot}^{\prime})^2, \mu_{\bot}^{\prime 2}\right), 
  \label{EvolEq}
\end{multline}
where $a,b$- are the flavour indices,  $x$- the fraction of the proton's longitudinal momentum  carried by the parton $a$, $k_{\bot}$-the transverse momentum, $\mu$ - the evolution scale, $\mu_0$- the initial evolution scale, $z$ - the  momentum transfer in the splitting and $z_M$- the soft gluon resolution scale which can be scale dependent. 
To treat the virtual/non-resolvable emissions, a new TMD Sudakov form factor was introduced \cite{Hautmann:2022xuc}
\begin{equation}
\Delta_a(\mu^2,\mu_0^2,k_{\bot}^2)\equiv\Delta_a(\mu^2,k_{\bot}^2)=\exp\left(-\sum_b\int_{\mu_0^2}^{\mu^2}\frac{d\mu'^2}{\mu'^2}\int_0^{z_M}dz\ z\bar P^R_{ba}(z,k_{\bot}^2,\mu'^2)\right), 
\label{TMDSud}
\end{equation}
using the angular averaged TMD splitting functions $\bar P^R_{ba}(z,k_{\bot}^2,\mu'^2)$. This construction was possible thanks to the momentum sum rule and unitarity. 
As an intermediate step,  a scenario with the TMD splittings included in the real resolvable emissions but with
the default PB Sudakov form factor 
\begin{equation}
\Delta_a(\mu^2,\mu_0^2)\equiv\Delta_a(\mu^2)=\exp\left(-\sum_b\int_{\mu_0^2}^{\mu^2}\frac{d\mu'^2}{\mu'^2}\int_0^{z_M}dz\ z P^R_{ba}(z,\mu^{\prime 2})\right)
\label{CollSud}
\end{equation}
was studied. 
It was shown analytically \cite{Hautmann:2022xuc}, that only when the same type of splitting functions are used both in the real emissions and Sudakov form factors, the evolution equation from Eq.~\ref{EvolEq} satisfies the momentum sum rule. 
In other words, for the evolution equation Eq.~\ref{EvolEq} with the TMD Sudakov form factor in the form given by Eq.~\ref{TMDSud} the momentum sum rule holds, whereas with the collinear Sudakov form factor from Eq.~\ref{CollSud} it is broken.

\begin{figure}[htb]
\begin{minipage}{0.49\textwidth}
\includegraphics[width=5.0cm]{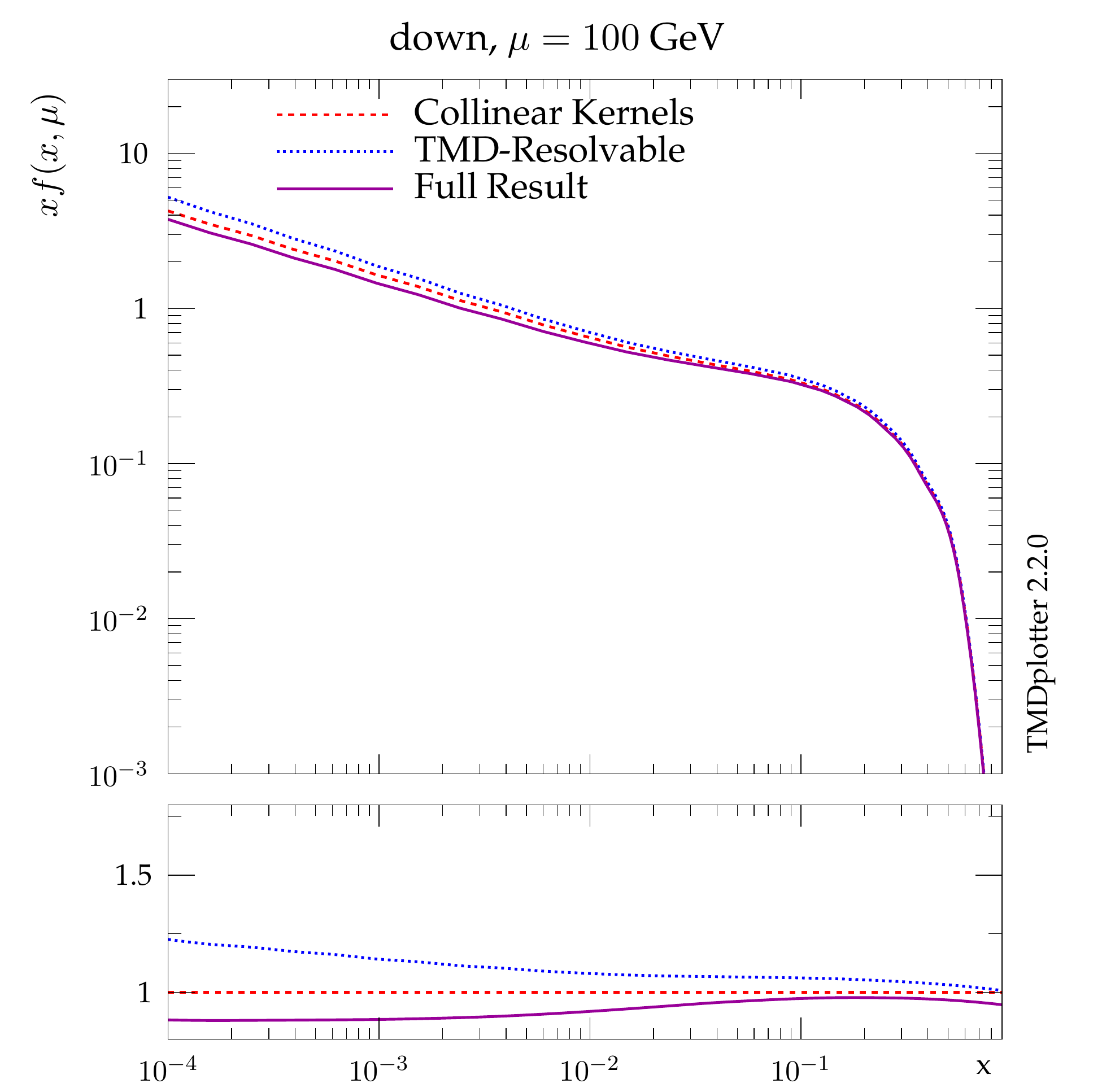}
\end{minipage}
\hfill
\begin{minipage}{0.49\textwidth}
\includegraphics[width=5.0cm]{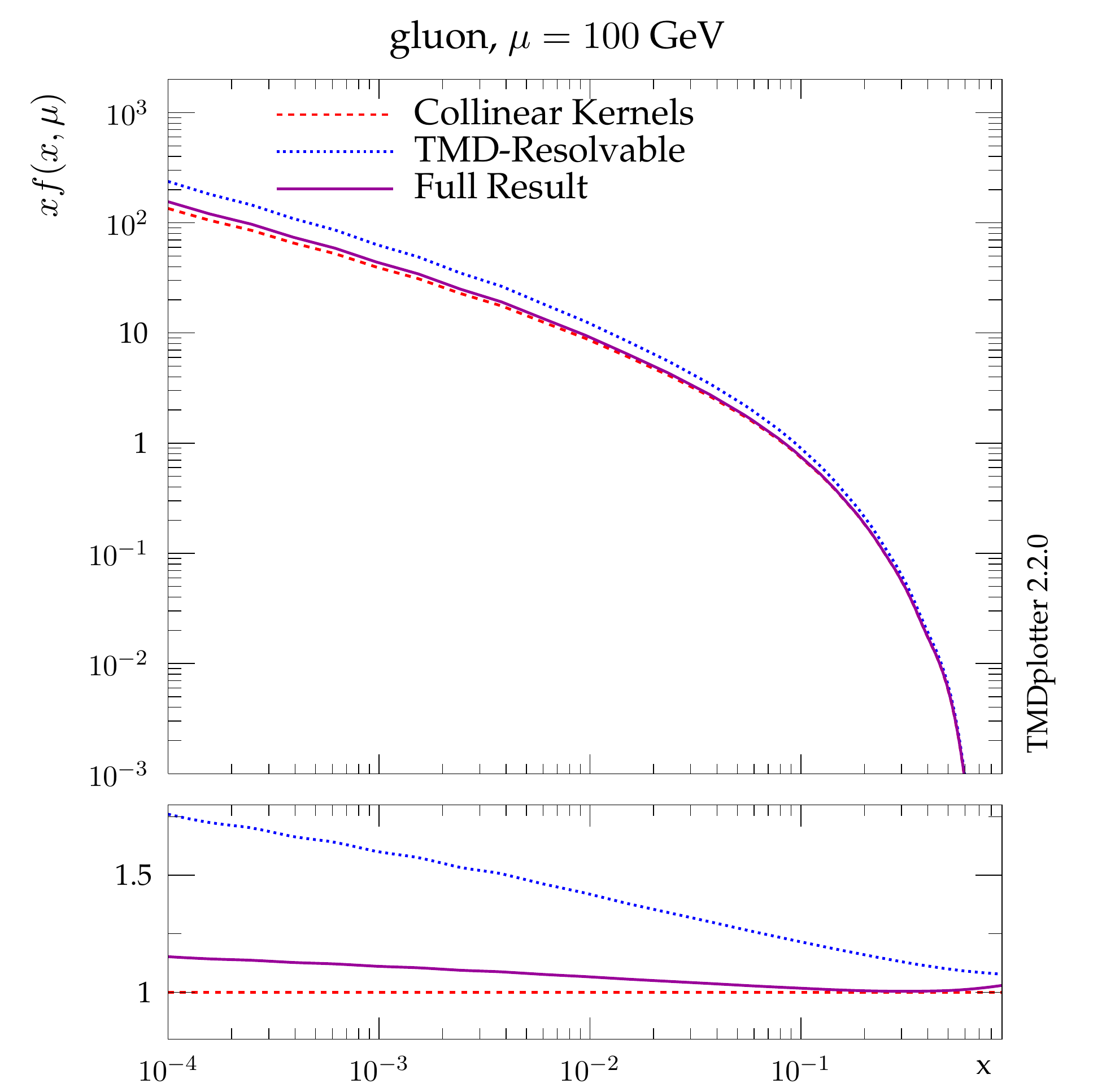}
\end{minipage}
\hfill 
\begin{minipage}{0.49\textwidth}
\includegraphics[width=5.0cm]{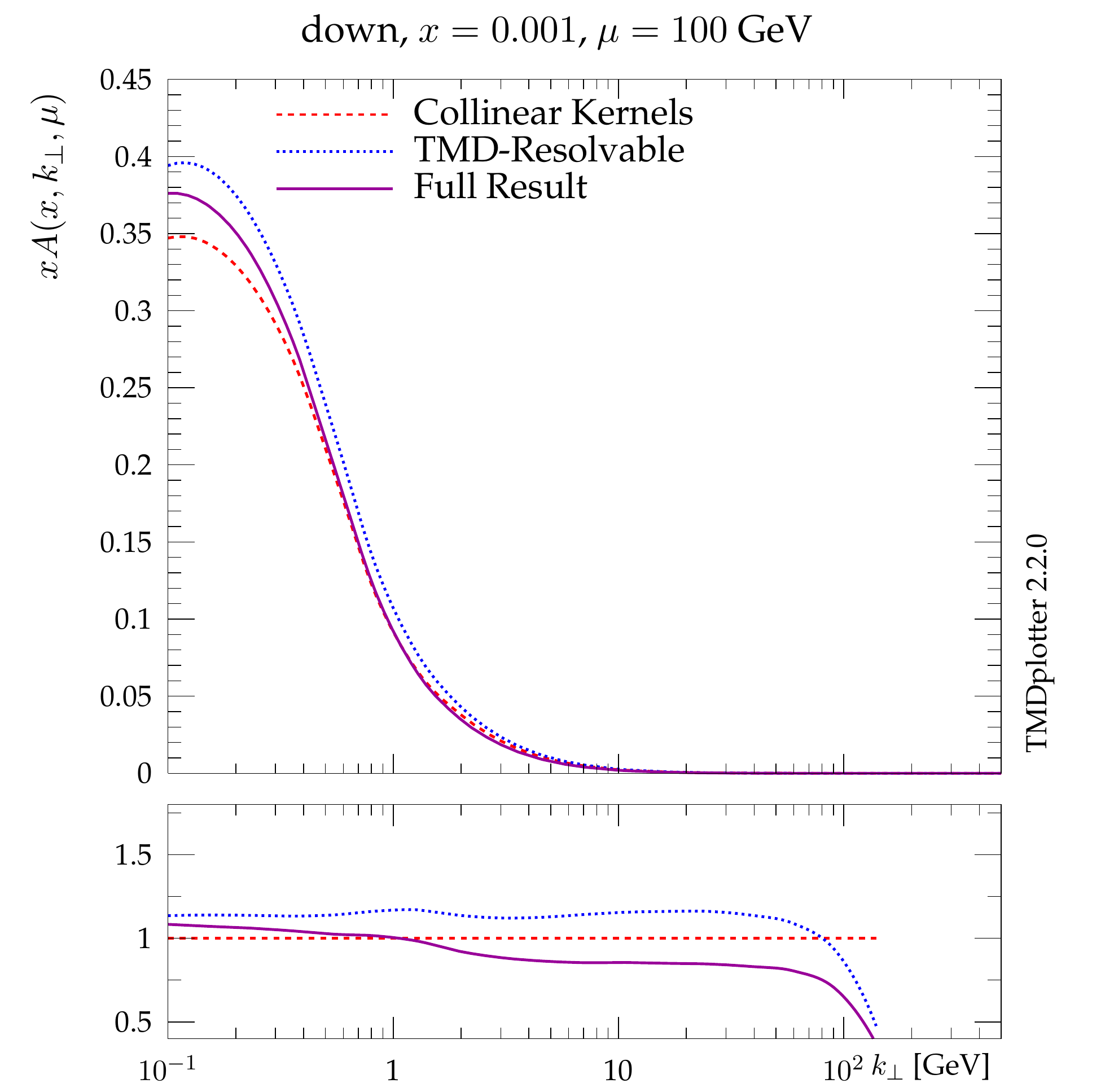}
\end{minipage}
\hfill 
\begin{minipage}{0.49\textwidth}
\includegraphics[width=5.0cm]{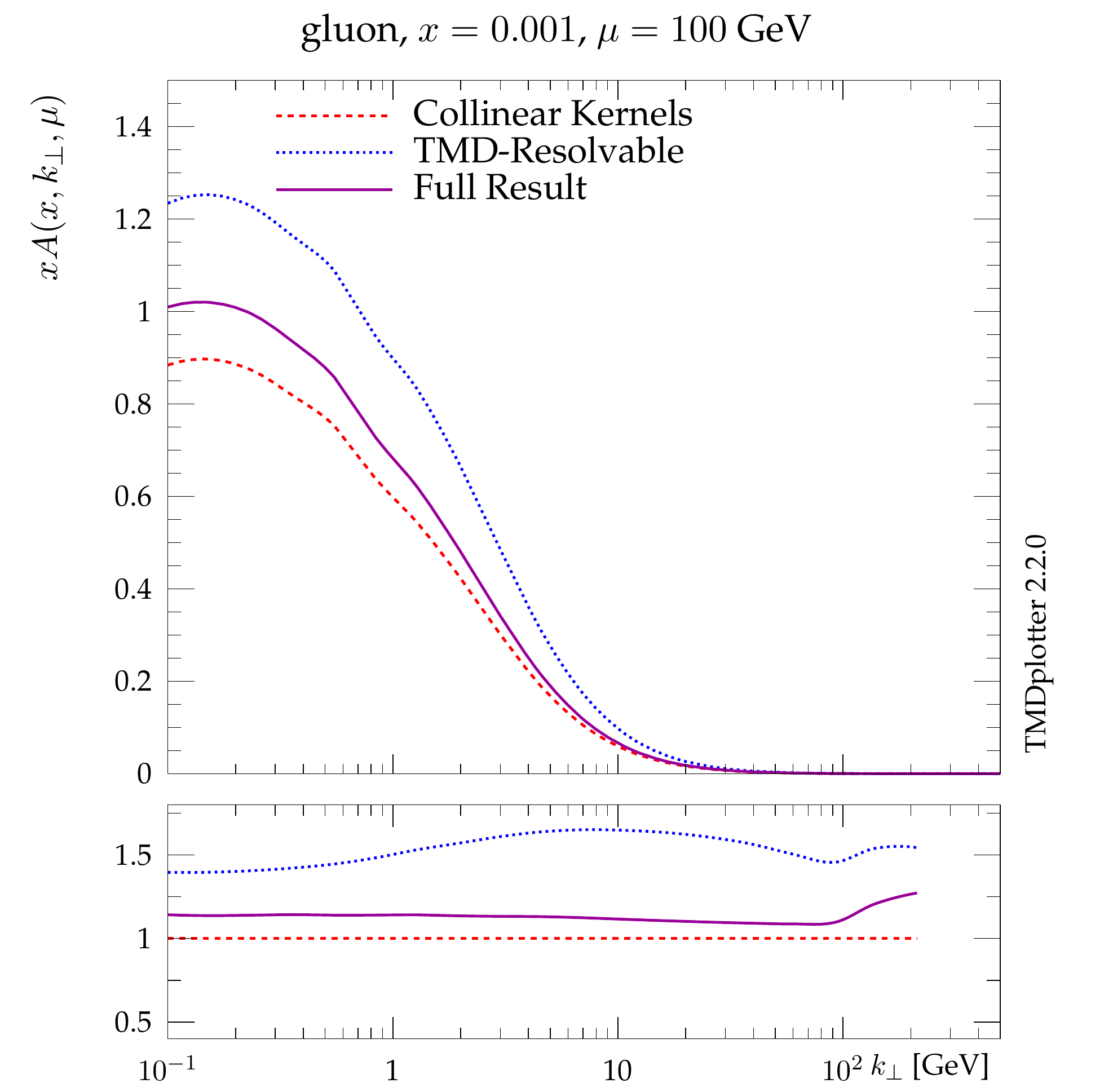}
\end{minipage}
\hfill 
\caption[]{
Down quark and gluon distributions for scenarios with the collinear splitting functions (red), with the TMD splitting functions in the real emissions and the collinear Sudakov form factor (blue) and with the TMD splitting functions both in the real emissions and in the Sudakov form factor (purple).
Top: integrated TMDs as a function of $x$ at $\mu=100\;\textrm{GeV}$. Bottom: TMDs as a function of $|k_{\bot}|$ at $x=0.001$ and $\mu=100\;\textrm{GeV}$ \cite{Hautmann:2022xuc}.  }
\label{Fig:Distributions}
\end{figure}

\section{Numerical results}
In the upper part of  Fig.~\ref{Fig:Distributions}, the integrated distributions (iTMDs) as a function of $x$ at the scale $\mu=100\;\textrm{GeV}$ are shown for down quark and gluon for 3 evolution scenarios: the dashed red curve is obtained from the  PB evolution equation with collinear splitting functions, the blue dotted curve with the model with TMD splitting functions in real resolvable emissions but with the collinear Sudakov form factors and the solid magenta line with the TMD splitting functions both in the real resolvable emissions and the Sudakov form factors.  In the bottom of Fig.~\ref{Fig:Distributions} the down quark and gluon TMDs as a function of $|k_{\bot}|$ are shown at $x=0.001$, $\mu=100\;\textrm{GeV}$ for the same 3 models. 
The bottom panel of each plot shows the ratios obtained with respect to the fully collinear scenario. 
For the purpose of this study, the same starting distribution was used for all those 3 models, which means that the differences between the curves come only from the evolution, i.e.  purely from the treatment of the splitting functions. For the iTMDs, the effect of the TMD splitting functions is visible especially at low $x$, for the TMDs, the effects are visible in the whole $k_{\bot}$ region. It is worth reminding that for both  the red and magenta curves the momentum sum rule holds, whereas the blue curve violates it. The numerical check of the momentum sum rule was performed in \cite{Hautmann:2022xuc}.

\section{Conclusions}
In this work a parton branching algorithm  to obtain TMDs and integrated distributions, which for the first time includes TMD  splitting functions and fulfils momentum sum rule, was presented. 
A new TMD Sudakov form factor was constructed using the momentum sum rule and unitarity.
The	studies presented here are at the level of the forward evolution but it is a 
first step towards a full TMD MC  generator covering the small-$x$ phase space.

\section*{Acknowledgements}  
Presented results were obtained  in collaboration with F. Hautmann, M. Hentschinski, L. Keersmaekers, A. Kusina and K. Kutak. 
A. Lelek acknowledges funding by Research Foundation-Flanders (FWO) (application number: 1272421N).

 \bibliographystyle{mybibstyle} 

    {

\bibliography{DiffractionLowx.bib}

    }

\end{document}